\documentclass[iop,apjl]{emulateapj}
\usepackage{amsfonts}
\usepackage[dvips]{color}
\usepackage{multirow}
\newcommand{\HII}{H\,{\sc ii}}

\newcommand{\herschel}{{\it Herschel }}

\newcommand{\OIII}{[O\,{\sc iii}]}
\newcommand{\NII}{[N\,{\sc ii}]}
\newcommand{\NIII}{[N\,{\sc iii}]}

\newcommand{\LNII}{$L_{\rm [NII]}$}
\newcommand{\LOIII}{$L_{\rm [OIII]}$}
\newcommand{\LIR}{$L_{\rm IR}$}

\newcommand{\NIIIZERO}{[N\,{\sc ii}] 205 $\mu$m}
\newcommand{\NIITWOONE}{[N\,{\sc ii}] 122 $\mu$m}
\newcommand{\NIIIO}{[N\,{\sc ii}]\,(1$\rightarrow$0)}
\begin{document}
\shorttitle{\NII\ 205 $\mu$\lowercase{m} line as a SFR indicator}
\shortauthors{Zhao et al.}
\title{A \herschel Survey of the \NII\ 205 $\mu$\lowercase{m} Line in Local Luminous Infrared Galaxies --- The \NII\ 205 $\mu$\lowercase{m} Emission as a Star Formation Rate Indicator$^\star$} 
\author{Yinghe Zhao\altaffilmark{1, 2, 3}, Nanyao Lu\altaffilmark{2}, C. Kevin Xu\altaffilmark{2}, Yu Gao\altaffilmark{1, 3}, S. Lord\altaffilmark{2}, J. Howell\altaffilmark{2}, K. G. Isaak\altaffilmark{4}, V.~Charmandaris\altaffilmark{5, 6}, T.~Diaz-Santos\altaffilmark{7}, P. Appleton\altaffilmark{2}, A. Evans\altaffilmark{8, 9}, K. Iwasawa\altaffilmark{10}, J.~Leech\altaffilmark{11}, J. Mazzarella\altaffilmark{2}, A.~O.~Petric\altaffilmark{12}, D.~ B.~ Sanders\altaffilmark{13}, B.~Schulz\altaffilmark{2}, J.~Surace\altaffilmark{7}, P.~P.~van~der~Werf\altaffilmark{14}}
\altaffiltext{1}{Purple Mountain Observatory, Chinese Academy of Sciences, Nanjing 210008, China; yhzhao@ipac.caltech.edu}
\altaffiltext{2}{Infrared Processing and Analysis Center, California Institute of Technology, MS 100-22, Pasadena, CA 91125, USA}
\altaffiltext{3}{Key Laboratory of Radio Astronomy, Chinese Academy of Sciences, Nanjing 210008, China}
\altaffiltext{4}{ESA Astrophysics Missions Division, ESTEC, PO Box 299, 2200 AG Noordwijk, The Netherlands}
\altaffiltext{5}{Department of Physics and ITCP, University of Crete, GR-71003 Heraklion, Greece}
\altaffiltext{6}{IESL/Foundation for Research and Technology - Hellas,  GR-71110, Heraklion, Greece and Chercheur Associ\'e, Observatoire de Paris, F-75014, Paris, France}
\altaffiltext{7}{Spitzer Science Center, California Institute of Technology, MS 220-6, Pasadena, CA 91125, USA}
\altaffiltext{8}{Department of Astronomy, University of Virginia, 530 McCormick Road, Charlottesville, VA 22904, USA}
\altaffiltext{9}{National Radio Astronomy Observatory, 520 Edgemont Road, Charlottesville, VA 22903, USA}
\altaffiltext{10}{ICREA and Institut de Ci\`encies del Cosmos (ICC), Universitat de Barcelona (IEEC-UB), Mart\'i i Franqu\`es 1, 08028, Barcelona, Spain}
\altaffiltext{11}{Department of Physics, University of Oxford, Denys Wilkinson Building, Keble Road, Oxford, OX1 3RH, UK}
\altaffiltext{12}{Astronomy Department, California Institute of Technology, Pasadena, CA 91125, USA}
\altaffiltext{13}{University of Hawaii, Institute for Astronomy, 2680 Woodlawn Drive, Honolulu, HI 96822, USA}
\altaffiltext{14}{Leiden Observatory, Leiden University, PO Box 9513, 2300 RA Leiden, The Netherlands}
\altaffiltext{$\star$}{\herschel is an ESA space observatory with science instruments provided by European-led Principal Investigator consortia and with important participation from NASA.}
\begin{abstract}
We present, for the first time, a statistical study of \NIIIZERO\ line emission for a large sample of local luminous infrared galaxies using {\it Herschel} Spectral and Photometric Imaging Receiver Fourier Transform Spectrometer (SPIRE FTS) data. For our sample of galaxies, we investigate the correlation between the \NII\ luminosity (\LNII) and the total infrared luminosity (\LIR), as well as the dependence of \LNII/\LIR\ ratio on \LIR, far infrared colors ({\it IRAS} $f_{60}/f_{100}$) and the \OIII\ 88 $\mu$m to \NII\ luminosity ratio. We find that \LNII\ correlates almost linearly with \LIR\ for non AGN galaxies (all having $L_{\rm IR} < 10^{12}\, L_\odot$) in our sample, which implies that \LNII\ can serve as a SFR tracer which is particularly useful for high redshift galaxies which will be observed with forthcoming submm spectroscopic facilities such as the Atacama Large Millimeter/submillimeter Array. Our analysis shows that the deviation from the mean \LNII-\LIR\ relation correlates with tracers of the ionization parameter, which suggests the scatter in this relation is mainly due to the variations in the hardness, and/or ionization parameter, of the ambient galactic UV field among the sources in our sample.
\end{abstract}

\section{Introduction}

The star formation rate (SFR) is one of the fundamental physical parameters for characterizing galaxies. Having an effective way to derive SFRs for galaxies spanning a large range of look-back times is crucial for our understanding of galaxy evolution leading back to initial conditions. Currently, SFRs have been inferred using continuum or spectral line emission from a wide range of wavelengths (e.g. Calzetti et al. 2009 and references therein; Kennicutt \& Evans 2012). However, few of these SFR indicators can be observed over a large redshift range using current facilities. Using the capabilities of the ESA {\it Herschel Space Observatory} (hereafter {\it Herschel}; Pilbratt et~al. 2010), it is possible to define new SFR indicators in the far infrared (FIR) window, which can be followed up by the Atacama Large Millimeter/submillimeter Array (ALMA) over a large redshift range.

The \NII\ 205 $\mu$m line, which arises from the $^3$P$_1$$\rightarrow$$^3$P$_0$ transition (205.197 $\mu$m; hereafter \NIIIO) of the ground state of the singly ionized nitrogen, is expected to be an excellent indicator of SFR. This FIR line emission essentially traces all of the warm ionized interstellar medium (ISM). The ionization potential of nitrogen (14.53 eV) is only $\sim 0.9$ eV larger than that of hydrogen (13.6 eV), allowing the \NII\ luminosity to be used as an estimate of the flux of ionizing photons (e.g. Bennett et al. 1994). The critical density of \NIIIO\ is only 44~cm$^{-3}$ at $T=8000$ K (Oberst et al. 2006), and the energy level that emits the line corresponds to only 70 K, so that it can be easily collisionally excited. In addition, the \NIIIZERO\ emission is usually optically thin because of its small Einstein coefficient, and suffers much less dust extinction than optical and near infrared lines. In particular, this line is accessible using ALMA for a significant fraction of the $0.6\lesssim z\lesssim16$ redshift range (bands 3-10){\footnotemark\footnotetext{Redshift intervals corresponding to different ALMA receivers: $11.6 \lesssim z\lesssim16.3$ (Band 3); $3.0\lesssim z\lesssim10.6$ (Bands $4-7$); $2.0\lesssim z\lesssim2.7$ (Band 8); $1.0\lesssim z\lesssim1.4$ (Band 9); $0.6\lesssim z\lesssim0.8$ (Band 10)}}. For more local galaxies, the \NIIIO\ line can be observed in the two windows around 200~$\mu$m ($1.25 -1.4$ THz and $1.45 -1.6$ THz; e.g. Paine et al. 2000; Yang et al. 2010) from the very driest sites with future facilities such as the Sub-mm/THz telescope at the Chinese Antarctic Observatory under development (e.g. Cui 2010). 

Early studies have shown that the \NIIIZERO\ line is comparatively bright. In most normal star-forming galaxies, it is the third to fifth brightest FIR/submm line after [C\,{\sc ii}] 158 $\mu$m, [O\,{\sc iii}] 88 $\mu$m, [O\,{\sc i}] 63 $\mu$m and \NIITWOONE\ lines (e.g., Wright et al. 1991; Brauher et al. 2008). The luminosity of \NIIIZERO\ line (\LNII) may be up to $\sim10^{-4}$ times the total infrared luminosity ($L_{\rm IR}$). For example, \LNII\ is about $3\times10^{-4}L_{\rm IR}$ and $5\times10^{-5}L_{\rm IR}$ in the Milky Way (Wright et al. 1991) and M82 (Petuchowski et al. 1994), respectively. With such a high luminosity, this line has already been detected with ALMA for luminous/ultraluminous infrared galaxies at high redshift (e.g. Nagao et al 2012; $z=4.76$).

The \NIIIZERO\ line is generally inaccessible to ground-based facilities, and its rest wavelength is longer than the cutoff of the Infrared Space Observatory ({\it ISO}) Long Wavelength Spectrometer (LWS). Therefore, this line was observed in only a handful of extragalactic objects using satellite and airborne platforms (e.g. Wright et al. 1991; Petuchowski et al. 1994; Lord et al. 1995) prior to the advent of {\it Herschel}. Here we report our first results on the \NIIIZERO\ line emission for a large sample of galaxies, observed with the Fourier-transform spectrometer (FTS) of the SPIRE instrument (Griffin et al. 2010) on board {\it Herschel}. We present the correlation between \LNII\ and \LIR, to characterize the \NIIIZERO\ line as a SFR indicator. We also investigate possible dependences of \NIIIZERO\ to IR luminosity ratio (\LNII/\LIR) on \LIR, FIR color ($f_{60}/f_{100}$, where $f_{60}$ and $f_{100}$ are the {\it IRAS} 60 $\mu$m and 100 $\mu$m fluxes, respectively), and the \OIII\ 88 $\mu$m to \NIIIZERO\ emission ratio (\LOIII/\LNII), to reveal the cause of the scatter in this relation. Throughout the paper, we adopt a Hubble constant of $H_0=70~$km~s$^{-1}$~Mpc$^{-1}$, $\Omega_{\rm M} =0.3$, and $\Omega_\Lambda=0.7$.

\section{Sample, Observations and Data Reduction}
The sample discussed in this {\it Letter} is part of the \herschel open time project {\it Herschel Spectroscopic Survey of Warm Molecular Gas in Local Luminous Infrared Galaxies (LIRGs)} (PI: N. Lu), which aims primarily at studying the dense and warm molecular gas properties of 125 LIRGs ($L_{\rm IR} \equiv L(8-1000\,\mu{\rm m)} >10^{11} L_\odot$, where $L_{\rm IR}$ was calculated by using the {\it IRAS} four-band fluxes and the equation given in Sanders \& Mirabel (1996)), which comprise a flux limited subset of the Great Observatories All-Sky LIRGs Survey sample (GOALS; Armus et al. 2009). The full program details as well as observational data on individual galaxies will be given in a future paper upon the expected completion of the program in early 2013. Here we present a subsample of 70 galaxies, including 63 LIRGs and 7 ULIRGs ($L_{\rm IR} > 10^{12}$ $L_\odot$), all of which are point sources with respect to the {\it Herschel} SPIRE/FTS beam. The determination of these galaxies as relative point sources was based both on their observed mid-infrared emission extent (for example, see D\'{i}az-Santos et al. 2010, 2011), and on the consistency between the observed emission strength seen in the overlapping FTS bands between 316 and 324 $\mu$m (beam sizes are $\sim 20''$ and $\sim 37''$ for SSW and SLW, respectively, in this wavelength range). A companion letter (Lu et al. 2013) gives a brief program description and some initial results on the observed CO line emission.

The observations were conducted with the SPIRE FTS in staring and high spectral resolution mode between 194--672 $\mu$m, yielding a resolution of 0.04 cm$^{-1}$ (=1.2 GHz). The data were reduced using a customized version of the standard SPIRE reduction and calibration pipeline for point source mode included in HIPE version 7.3 (Ott 2010). We used an averaged telescope relative spectral response function (RSRF) in place of the daily RSRF, to perform the flux calibration. During the reduction, we did not apply any apodization to the spectra. To extract the final spectrum, a ``background" emission, which was obtained by fitting the median spectrum from the spatially surrounding detectors, was subtracted from the central detector.

\begin{figure}
\centering
\includegraphics[width=.4\textwidth,bb=26 15 644 826]{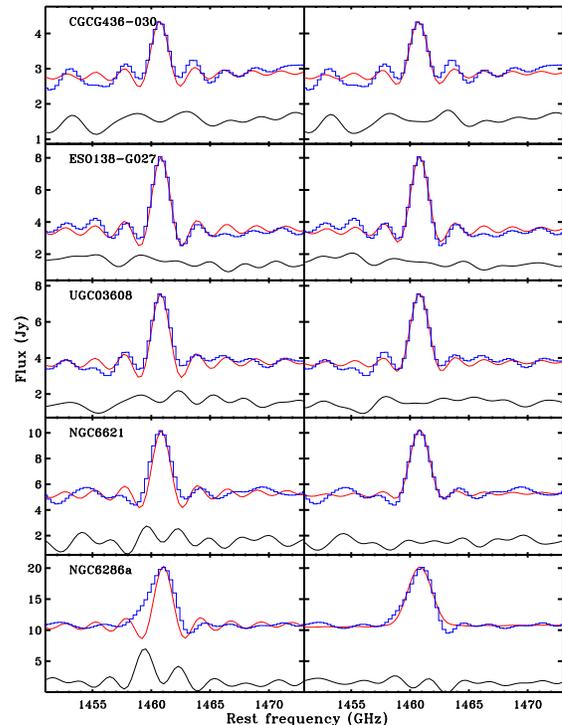}
\caption{Typical spectra (corrected to the source rest frame) of the \NIIIZERO\ line for galaxies with a range of $f_{60}/f_{100}$ colors (going redder from top to bottom). In each panel, the blue, red and black lines show the observed spectrum, the fitted profile of the \NII\ line plus continuum, and the residual, respectively. In the left panels, the \NII\ line is fitted using a {\it Sinc} function with a free width parameter. The right panels show the same data but the line is fitted using a fixed instrumental {\it Sinc} function convolved with a free width gaussian shape (SCG). It is clear that the SCG function can give a better fitting result for this line.}
\label{Fig1}
\end{figure}

In most cases the \NIIIZERO\ line is the brightest line in the SPIRE FTS wavelength range, and almost all sources show high signal-to-noise ratio (S/N) in the \NII\ line. In Fig. 1 we plot several typical spectra of the \NII\ line for galaxies with a range of FIR colors ($0.4\lesssim f_{60}/f_{100}\lesssim1.4$), and superimpose a fitted continuum and the line profile. From the figure, one can see that the fixed {\it Sinc} function convolved with a free width gaussian (SCG) profile gives a better fitting result for the \NII\ line since the instrumental resolution of the SPIRE FTS in high resolution mode at $\sim 210$ $\mu$m is only about 300 km s$^{-1}$ and the line width has an influence on the line shape. Therefore, we adopted the SCG profiles for the integrated \NII\ line flux measurements except for one galaxy (CGCG448-020) where a single {\it Sinc} profile was better due to its narrow line width ($<200$ km s$^{-1}$ for low J CO lines; e.g. Baan et al. 2008; Leech et al. 2010) and relatively low S/N ($\sim 4$) in the line. The resulting full width at half maximum of the \NII\ line (excluding CGCG448-020) is $140-610$ km s$^{-1}$, with a median value of  $\sim 300$ km s$^{-1}$. In most cases, the $1\sigma$ statistical uncertainties in the integrated line flux are less than 10\%, and the median value is 7\%.

\section{Results and Discussions}
\subsection{\LNII-$L_{\rm IR}$ Correlation}

\begin{figure*}[t]
\centering
\includegraphics[width=0.68\textwidth,bb=80 243 550 666]{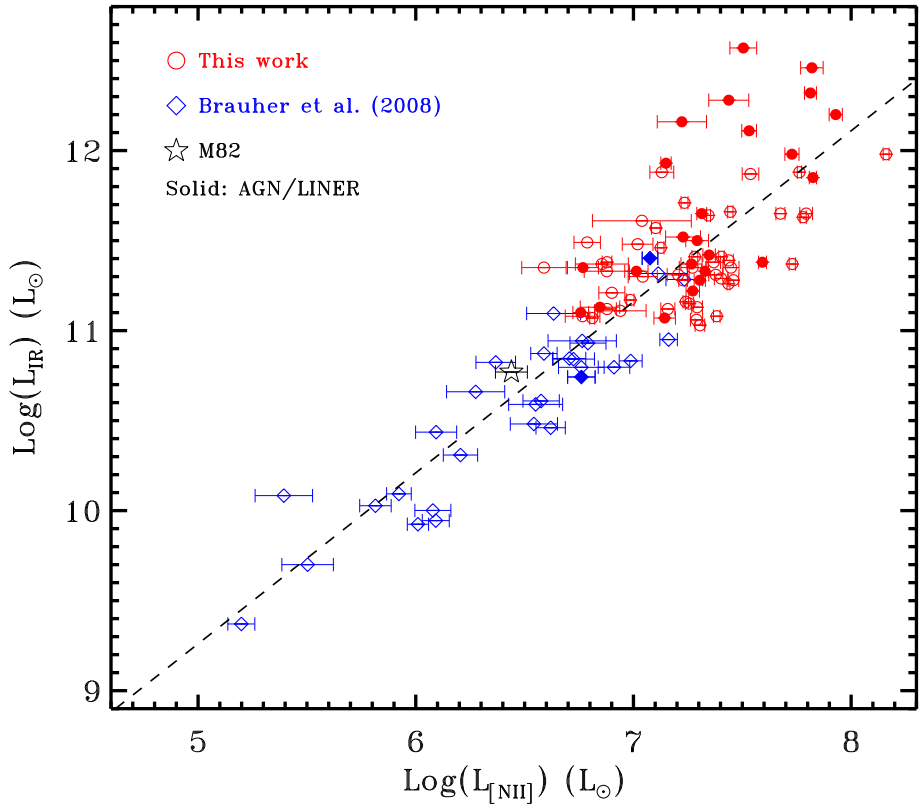}
\caption{\NIIIZERO\ plotted against total infrared luminosities for normal and luminous infrared galaxies. Circles (red) are galaxies having {\it Herschel} observations. Diamonds (blue) are galaxies from Brauher et al. (2008), for which an additional uncertainty of $\sim 0.3$ dex should be taken into account besides the plotted error bar. The superposed line represents a geometrical mean, least-squares linear fit to all data points except for those solid symbols.}
\label{Fig2}
\end{figure*}

Our sample galaxies are all (U)LIRGs and 23 out of 70 show AGN/LINER features (identified using optical, mid infrared, or X-ray data). In order to increase the size and dynamic range of the sample used to study the \LNII-$L_{\rm IR}$ relation, we also include 30 unresolved galaxies observed by {\it ISO} LWS and compiled by Brauher et al. (2008), in our analysis. For these objects, most of which are normal star-forming galaxies, the luminosity distances were derived with the same method as used for our (U)LIRG sample (e.g. Armus et al. 2009), i.e. by correcting the heliocentric velocity for the 3-attractor flow model of Mould et al. (2000). Their \NIIIZERO\ fluxes are derived from the observed \NIITWOONE\ fluxes given in Brauher et al. (2008), using the theoretical \NIITWOONE\ to \NIIIZERO\ emission ratio (hereafter $R_{21}$) at $n_e=80$ cm$^{-3}$ (the median value for late type galaxies; see e.g. Ho et al. 1997) of $R_{21}$=2.6. Here we compute the theoretical $R_{21}$ by assuming electron impact excitation for the ground-state levels of N$^+$, and using the fitted results (Draine 2011) of Hudson \& Bell (2005) with $T=8000$ K to calculate the collision strengths (more details are provided in Zhao et al. 2013, in preparation). Regarding the inclusion of Brauher et al. galaxies in our study, we note that (1) about 90\% of the galaxies in Ho et al. (1997) have $n_e<300$ cm$^{-3}$, yielding a theoretical $R_{21}= 0.94-5.3$ ($n_e=10-300$ cm$^{-3}$); (2) empirically, we see an observed $R_{21}$ of $1.5-4.3$ for the nine galaxies (eight in our sample plus M82) that have both \NIITWOONE\ and \NIIIZERO\ data. Therefore, the uncertainty in \LNII\ caused by extrapolating \NIIIZERO\ fluxes from \NIITWOONE\ emissions is a factor of a few ($\sim 0.3$ dex), and is significantly smaller than the two order-of-magnitude span of \LIR\ found in our sample, thus yielding useful results concerning the \LNII-\LIR\ relation.
 
In Fig. 2, we plot \LNII\ against \LIR\ for both our (U)LIRGs (cirles) and Brauher et al.'s galaxies (diamonds), as well as the starburst galaxy M82 (star-symbol; Petuchowski et al. 1994). The solid symbols indicate AGNs/LINERs. Generally, \LNII\ increases with \LIR, albeit the scatter seems to increase with \LIR\ (we will return to this later). In fact, the scatter at the high \LIR\ end will be much reduced if we only look at the non-active galaxies. A nonweighted least-squares linear fit, using a geometrical mean functional relationship (Isobe et al. 1990), to the non-active galaxies, gives
\begin{equation}
\log L_{\rm IR}=(4.51\pm0.32)+(0.95\pm0.05)\log L_{\rm[NII]}
\end{equation}
with a scatter of 0.26 dex in \LIR. The Spearman rank correlation coefficient $\rho$ of the trend is 0.78 at a $>5\sigma$ level of significance, which indicates a very strong correlation between \LNII\ and \LIR. Our fit shows that, on average, \LIR\ scales with $L_{\rm [{N\,II]}}^\alpha$ with $\alpha$ between 0.8 and 1.1 at the $3\sigma$ significance. This nearly linear relation suggests that the power source of the \NIIIZERO\ emission tightly relates to the star-forming activities. Using the SFR-\LIR\ calibration given in Kennicutt \& Evans (2012), we can also build a direct relation between SFR and \LNII, namely
\begin{equation}
\log {\rm SFR}\,(M_\odot\,{\rm yr}^{-1})=(-5.31\pm0.32)+(0.95\pm0.05)\log L_{\rm[NII]}\,(L_\odot)
\end{equation}

Comparing to the IR luminosity, the \NIIIZERO\ emission could be a more convenient SFR indicator in the sense that it is more unambiguous, i.e. less affected by emissions from older stars (e.g. in the Milkyway the \NIIIZERO\ flux is lower than that predicted by a linear \LNII-\LIR\ relation when IR intensity going lower; see Fig. 4 in Bennett et al. 1994), and inter-comparable since the techniques for obtaining \LIR\ for low and high redshift objects are so instrument-dependent. The majority of massive galaxies up to $z = 5-6$ are already chemically evolved systems based on their CO and dust emissions (Solomon \& Vanden Bout 2005) or optical/IR line spectroscopy  (Hamann \& Ferland 1999). Therefore, Eq. (2) can be reasonably applicable to high-$z$ starburst galaxies despite that the \LNII-\LIR\ relation possibly depends on metallicity.

\subsection{The scatter in the \LIR-\LNII\ relation}
\begin{figure*}[t]
\centering
\includegraphics[width=0.88\textwidth,bb=0 100 680 552]{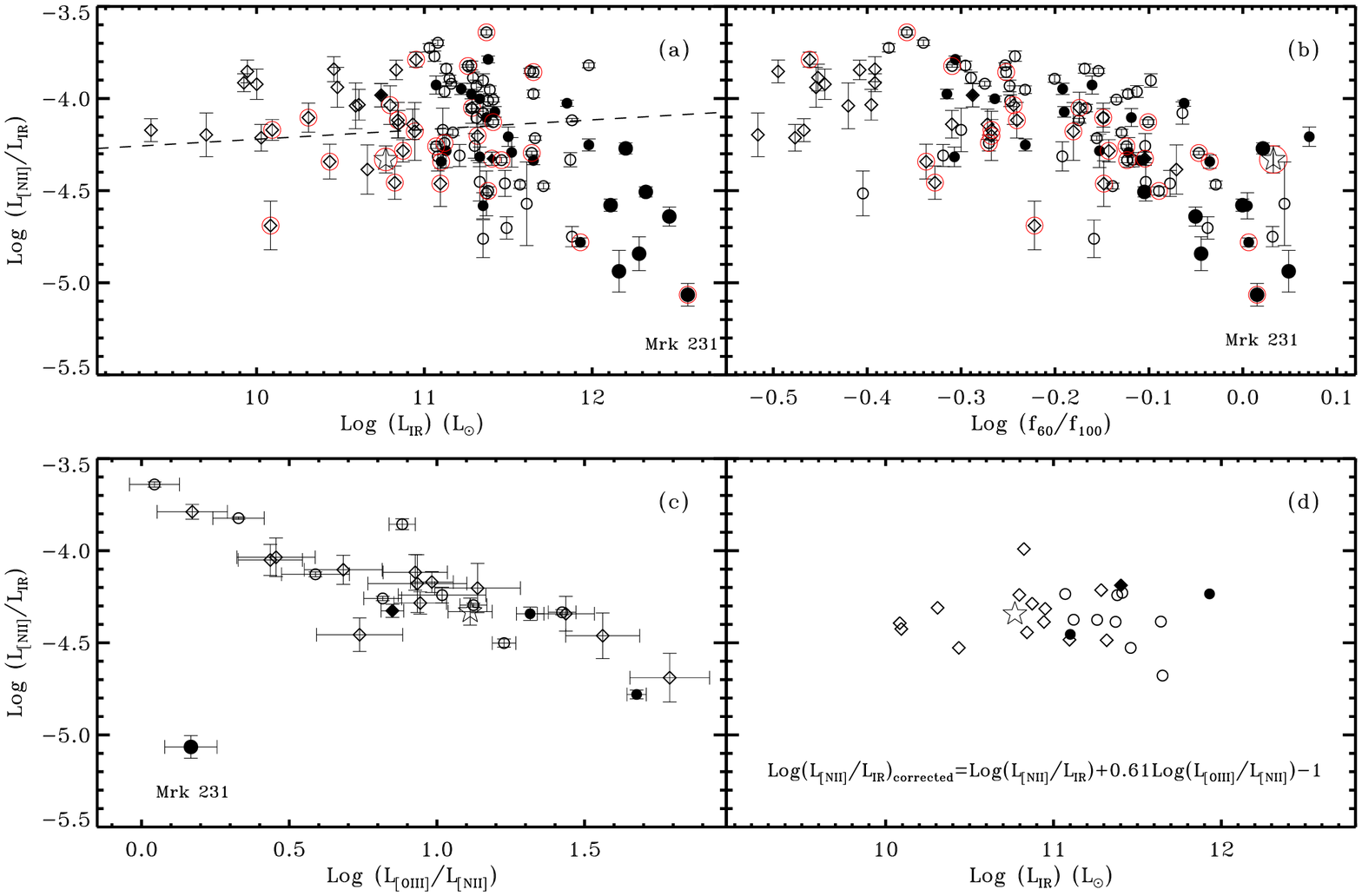}
\caption{Ratio of \NIIIZERO\ to IR luminosity is plotted against ({\it a}) and ({\it d}): IR luminosity, ({\it b}): {\it IRAS} 60 $\mu$m/100 $\mu$m color, and ({\it c}): [O\,{\sc iii}] 88 $\mu$m/\NIIIZERO\ flux ratio. Symbols are the same as Fig. 2 except that here larger circles represent ULIRGs. Symbols enclosed by an additional red circle in panels ({\it a}) and ({\it b}) are the same sources as shown in panels ({\it c}) and ({\it d}). Eq. (1) is shown by the dashed line in panel ({\it a}). The ordinate of panel ({\it d}) is a nominal value after the correction for the hardness effect (using the equation shown in this panel). Mrk 231 is outside the plotted range of panel ({\it d}).}
\label{Fig3}
\end{figure*}

In Fig. 2 we can also see an indication of a steeper relation between \LNII\ and \LIR\ for $L_{\rm IR} > 10^{11}\,L_\odot$ galaxies. This is more clearly demonstrated in Fig. 3{\it a}, in which we plot \LNII/\LIR\ vs \LIR\ for the same set of galaxies. For ULIRGs, the mean \LNII/\LIR\ ($-4.69$ dex) is considerably lower than that ($-4.14$ dex) of the galaxies with lower IR luminosities. Due to the small ULIRG sample size and the prevalence of AGN we can reach no firm conclusion as to whether the \LNII/\LIR\ decreases with \LIR\,$> 10^{12}\, L_\odot$ for starburst-dominated galaxies.

However, the scatter in the \LIR-\LNII\ relation is as large as $\sim 0.3$ dex, and seems to increase with \LIR, specially when ULIRGs are taken into account. The scatter could have several different origins: metallicity variations and most likely, differing ionizing conditions. These latter effects include: the stellar UV input via the initial mass function, the aging of starbursts, the \HII\ region/ISM geometry or the presence of an AGN. Our sample of galaxies contains few low metallicity objects, and therefore we only address ionization effects here.

We note that the ionization parameter ($U$) can be especially high in (U)LIRGs (e.g. Farrah et al. 2007, Petric et al. 2011), where $U$ is defined as the dimensionless ratio of the number density of incident ionizing photons to the number density of hydrogen nuclei. Clearly, $U$ can be measured through ionization parameter-sensitive line ratios, such as [O\,{\sc iii}]/[O\,{\sc ii}] and [N\,{\sc iii}]/[N\,{\sc ii}]. Moreover, theoretical models have also shown that $U$ can be roughly indicated by the IR color, $f_{60}/f_{100}$,  by the means that $\log U$ increases with $f_{60}/f_{100}$ (see e.g. Abel et al. 2009).

In Fig. 3{\it b} we plot the dependence of \LNII/\LIR\ on $f_{60}/f_{100}$. We can see from the figure that there only exists a very weak dependence of \LNII/\LIR\ on $f_{60}/f_{100}$ when $\log\,(f_{60}/f_{100}) < -0.2$. However, \LNII/\LIR\ clearly decreases as $f_{60}/f_{100}$ increases when $\log\,(f_{60}/f_{100}) \gtrsim -0.2$, except for a few galaxies with the warmest FIR colors in the sample. Moreover, the scatter in this correlation is less than that in the  \LNII/\LIR\ vs. \LIR\ plot (see Fig. 3{\it a}) and depends little on $f_{60}/f_{100}$. But we note that the scatter in the \LNII/\LIR-$f_{60}/f_{100}$ plot is still large, and we hypothesize that it is caused by the variations in the radiation field, since N$^+$ needs high energy (29.6 eV) to form N$^{++}$, and the population of ionic N is determined mainly by the slope of the background UV spectrum, i.e. the hardness, at a given $U$.

To try to account for the effects of the hardness of ionizing radiation field on \LNII/\LIR,  we use the ratio of the \OIII\ 88 $\mu$m emission, rather than \NIII\ 57 $\mu$m emission where there are few detections, to the \NIIIZERO\ emission to trace the hardness of the ionizing photons. It takes 35 eV to form O$^{++}$, which is only $\sim 5$ eV higher than that needed for the formation of N$^{++}$, and thus the \OIII/\NII\ line ratios are even stronger indicator of the hardness. Furthermore, this ratio is insensitive to the gas density (Rubin 1984). Except for M82 (Duffy et al. 1987) and Mrk 231 (Fischer et al. 2010), all of the \OIII\ 88 $\mu$m fluxes are adopted from Brauher et al. (2008).

Fig. 3{\it c} shows the relation between \LOIII/\LNII\ and \LNII/\LIR. Clearly, there exists a good correlation such as \LNII/\LIR\ decreases as \LOIII/\LNII\ increases. There is only one outlier, the ULIRG Mrk 231, which shows both small \LNII/\LIR\ and \LOIII/\LNII. Excluding Mrk 231, and using an ordinary least-squares linear fit to the data, we obtain $\log (L_{\rm [NII]}/L_{\rm IR})=(-3.63\pm0.06)+(-0.61\pm0.06)\log (L_{\rm [OIII]}/L_{\rm [NII]})$. The Spearman rank correlation coefficient $\rho$ of this correlation is $-0.93$ at a $>5\sigma$ level of significance.

Galaxies plotted in Fig. 3{\it c} (hereafter \OIII\ sample) are also marked with red circles in Figs. 3{\it a} and 3{\it b}, which indicates that the hardness variation of the heating radiation field among galaxies can largely account for the scatter shown in the \LIR-\LNII\ relation, at least for $L_{\rm IR} < 10^{12}$ $L_\odot$ galaxies. Actually, this is also supported by the fact that the scatter of less luminous galaxies (0.2 dex for Brauher et al. sample; thus probably less active in star formation and lower $U$) is smaller than that of our luminous sample (0.28 dex). Keeping in mind that the conversion from \NIITWOONE\ flux to \NIIIZERO\ flux may contribute largely to the scatter of Brauher et al. sample, the real scatter in the \LNII-\LIR\ relation might probably be even smaller for these galaxies.

To obtain a direct view of how much the scatter can be reduced, we also display the hardness-corrected \LNII/\LIR, which was calculated according to the fitted \LOIII/\LNII-\LNII/\LIR\ relation, for the \OIII\ sample in Fig. 3{\it d} (arbitrarily scaled). Comparing to Fig. 3{\it a}, one can see that the scatter is reduced significantly (from 0.26 dex to 0.11 dex by excluding Mrk 231, which is still an outlier due to that the hardness correction has little effect on it, and is outside the plotted range).

All of the seven ULIRGs in our sample show some deficit in \LNII\ relative to \LIR. Four of the seven, including Arp 220, and Mrk 231, were observed by {\it ISO}  LWS, but no \OIII~88 $\mu$m emission was detected at the 3$\sigma$ limit (e.g. Brauher et a. 2008), despite all showing some AGN characteristics. These galaxies also show very warm FIR colors (see Fig. 3{\it b}). Therefore, very dusty \HII\ regions with high $U$ might be needed to explain the deficit of \NII\ emission in these galaxies (e.g. Fischer et al. 1999; Abel et al. 2009; Draine 2011).

\section{Summary}
In this {\it Letter} we have presented our initial results of the \NIIIZERO\ emission for a sample of 70 (U)LIRGs observed by the {\it Herschel} SPIRE FTS. Combining with the {\it ISO} and {\it IRAS} data, we have studied the correlation between \LNII\ and \LIR, and investigated the dependence of \LNII/\LIR\ on \LIR, FIR color $f_{60}/f_{100}$ and the UV hardness indicator \OIII/\NII. The main conclusions of our work are:
\begin{enumerate}
\item
For star-forming galaxies with $L_{\rm IR} < 10^{12}\,L_\odot$, there exists a strong correlation between \LNII\ and \LIR, namely $\log L_{\rm IR}=(4.51\pm0.32)+(0.95\pm0.05)\log L_{\rm[NII]}$, implying that the \NII\ emission can be used as an effective measure on SFR, particularly for sources where \LIR\ is difficult to measure such as high-$z$ galaxies. 

\item
For galaxies with $L_{\rm IR} < 10^{12}\,L_\odot$ in our sample, the \LNII/\LIR\ ratio is in the range of $\sim -4.8$ dex and $\sim -3.6$ dex, and varies systematically with the hardness/ionization parameter of the radiation field. However, additional effects are needed to explain the scatter in the \LNII-\LIR\ relation and their relatively lower \LNII/\LIR\ ratio for more IR luminous galaxies in our sample.
\end{enumerate}


\acknowledgements
We thank an anonymous referee for his/her comments, which improved this manuscript. This work is based in part on observations made with {\it Herschel}, a European Space Agency Conerstone Mission with significant participation by NASA. Support for this work was provided in part by NASA through an award issued by JPL/Caltech. Research for this project is partially supported by the Natural Science Foundation of China (grant 10833006) and Jiangsu Province (SBK201120678). Y.Z. thanks IPAC for the hospitality and the financial support during his visit. This research has made use of the NASA/IPAC Extragalactic Database (NED), which is operated by the Jet Propulsion Laboratory, California Institute of Technology, under contract with the National Aeronautics and Space Administration.


\begin{thebibliography}{}
\bibitem[]{}Abel, N. P., Dudley, C., Fischer, J., Satyapal, S., \& van Hoof, P. A. M. 2009, \apj, 701, 1147
\bibitem[]{}Armus, L., Mazzarella, J. M., Evans, A. S., et al. 2009, \pasp, 121, 559
\bibitem[]{}Baan, W. A., Henkel, C., Loenen, A. F., Baudry, A., \& Wiklind, T. 2008, \aap, 477, 747
\bibitem[]{}Bennett, C. L., Fixsen, D. J., Hinshaw, G., et al. 1994, \apj, 434, 587
\bibitem[]{}Brauher, J. R., Dale, D. A., \& Helou, G. 2008, \apjs, 178, 280
\bibitem[]{}Calzetti, D., Sheth, K., Churchwell, E., \& Jackson, J., The Evolving ISM in the Milky Way and Nearby Galaxies, 2009, Eds.: K. Sheth, A. Noriega-Crespo, J. Ingalls, \& R. Paladini
\bibitem[]{}Cui, X. 2010, Highlights of Astronomy, 15, 639
\bibitem[]{}D\'{i}az-Santos, T., Charmandaris, V., Armus, L. et al. 2010, \apj, 723, 993
\bibitem[]{}D\'{i}az-Santos, T., Charmandaris, V., Armus, L. et al. 2011, \apj, 741, 32
\bibitem[]{}Draine, B. T. 2011, Physics of the Interstellar and Intergalactic Medium (Princeton and Oxford: Princeton University Press)
\bibitem[]{}Duffy, P. B., Erickson, E. F., Haas, M. R., \& Houck, J. R. 1987, \apj, 315, 68
\bibitem[]{}Farrah, D., Bernard-Salas, J., Spoon, H., et al. 2007, \apj, 667, 149
\bibitem[]{}Fischer, J., Luhman, M. L., Satyapal, S., et al. 1999, Ap\&SS, 266, 91
\bibitem[]{}Fischer, J., Sturm, E., Gonz\'alez-Alfonso, E., et al. 2010, A\&A, 518, L41
\bibitem[]{}Griffin, M. J., Abergel, A., Abreu, A., et al. 2010, A\&A, 518, L3
\bibitem[]{}Hamann, F., \& Ferland, G. 1999, \araa, 37, 487
\bibitem[]{}Ho, L. C., Filippenko, A. V., \& Sargent, W. L. W. 1997, \apj, 487, 579
\bibitem[]{}Hudson, C. E, \& Bell, K. L 2005, A\&A, 430, 725
\bibitem[]{}Isobe, T., Feigelson, E. D., Akritas, M. G., \& Babu, G. J. 1990, \apj, 364, 104
\bibitem[]{}Kennicutt, R. C., \& Evans, N. J. 2012, ARA\&A, 50, 531
\bibitem[]{}Leech, J., Isaak, K. G., Papadopoulos, P. P., Gao, Y., \& Davis, G. R. 2010, \mnras, 406, 1634
\bibitem[]{}Lu, N., Zhao, Y., Xu, C. K., et al. 2012, in preparation
\bibitem[]{}Lord, S. D., Hollenbach, D. J., Colgan, S. W. J., et al., in Astronomical Society of the Pacific, Airborne Astronomy Symposium on the Galactic Ecosystem: From Gas to Stars to Dust, 1995, Vol 73, p151-158
\bibitem[]{}Mould, J. R., Huchra, J. P., Freedman, W. L., et al. 2000, \apj, 529, 786
\bibitem[]{}Nagao, T., Maiolino, R., De Breuck, C., Caselli, P., Hatsukade, B., \& Saigo, K. 2012, A\&A, 542, L34
\bibitem[]{}Oberst, T. E., Parshley, S. C., Stacey, G. J., et al. 2006, \apj, 652, L125
\bibitem[]{}Ott, S. 2010, ASP Conference Series, 434, 139
\bibitem[]{}Paine, S., Blundell, R., Papa, D. C., Barrett, J. W., \& Radford, S. J. E. 2000, \pasp, 112, 108
\bibitem[]{}Petric, A. O., Armus, L., Howell, J., et al. 2011, \apj, 730, 28
\bibitem[]{}Petuchowski, S. J., Bennett, C. L., Haas, M. R., et al. 1994, \apj,  427, L17
\bibitem[]{}Pilbratt, G. L., Riedinger, J. R., Passvogel, T., et al. 2010, A\&A, 518, L1
\bibitem[]{}Rubin, R. H. 1984, \apjs, 57, 349
\bibitem[]{}Sanders, D. B., \& Mirabel, I. F. 1996, 34, 749
\bibitem[]{}Solomon, P. M., \& Vanden Bout, P. A. 2005, \araa, 43, 677
\bibitem[]{}Wright, E. L., Mather, J. C., Bennett, C. L., et al. 1991, \apj, 381, 200
\bibitem[]{}Yang, H., Kulesa, C. A., Walker, C. K. et al., 2010, \pasp, 122, 490
\bibitem[]{}Zhao, Y., Lu, N., Xu, C. K., et al. 2013, in preparation

\end{thebibliography}
\end{document}